# Nonlinear effects in the phonon system of diamond crystal

A.S. Naumovets[1], Yu.M. Poluektov[1,2], V.D. Khodusov[1]

[1]V.N. Karazin Kharkiv National University, 4 Svobody Sq., Kharkov, 61022, Ukraine

[2]National Science Center "Kharkov Institute of Physics and Technology", National Academy of Sciences of Ukraine, 1 Akademicheskaya St., Kharkov 61108, Ukraine

**Abstract**

Thermodynamic properties of diamond are theoretically investigated on the ground of self-consistent description of a phonon gas in lattice, which generalizes the Debye model with taking into account the phonon-phonon interaction. It is shown that, at high temperatures, the theory predicts the linear in the temperature deviation of the isochoric heat capacity from the Dulong-Petit law. Unlike for the most crystals, where the decrease in the isochoric heat capacity is observed, our calculations for diamond predict the linear increase of the isochoric heat capacity with the temperature, viewed experimentally. The isobaric heat capacity of diamond, similar to other substances, linearly increases with the temperature.

**Keywords**: phonons; thermal properties; phonon-phonon interaction; Debye energy; elastic moduli.

## 1. Introduction

The Debye model [1] is the simplest way to described thermodynamic properties of dielectric crystals. In this approximation a crystal is considered as a continuous elastic isotropic medium in which oscillations propagate with the average speed $c_D$. At temperatures of the order of the Debye temperature $\Theta_D$ and higher, the phonon density in a crystal becomes higher than the particle number density, so nonlinear effects get appeared due to phonon interactions. A generalization of the Debye model which takes into account the phonon-phonon interaction was proposed in [2,3].

In many cases properties of crystals of certain symmetry can be well approximated by a model of an isotropic continuous medium, if its elastic moduli are chosen optimally [4]. A similar approach for an approximate account of nonlinear effects, which are cubic in the strain tensor degree, was used in [5]. In this model, a nonlinear elastic medium is characterized by five elastic moduli. In many cases, for a more complete description of the nonlinear effects, it is necessary to take into account fourth-order terms by the strain tensor in the free energy as well. For these cases, the nonlinear elastic properties of the medium are characterized by nine elastic moduli [6–9].

The behavior of the heat capacity of diamond at high temperatures differs from the behavior of the heat capacity of most crystals of other symmetry. Here it is necessary to distinguish the heat capacity at a constant volume $C_V$ (isochoric) and the heat capacity at a constant pressure $C_p$ (isobaric). The isochoric heat capacity is a more fundamental characteristic [10], but the isobaric heat capacity is usually determined experimentally. In the high temperature limit in accordance to the Dulong - Petit law [1],

$$C_V = 3N\left[1 - \frac{1}{20}\left(\frac{\Theta_D}{T}\right)^2\right], \qquad (1)$$

the isochoric heat capacity tends to a constant value. The difference between the heat capacities in the high-temperature limit is proportional to the temperature [1]:

$$C_p - C_V = 9n^2\gamma_T \Gamma^2 VT, \qquad (2)$$

where $n = N/V$ is the particle number density; $\Gamma = \partial \ln \Theta / \partial \ln n$ is the Grüneisen parameter; $\gamma_T = n^{-1}(\partial n/\partial p)_T$ is the coefficient of isothermal compressibility. Thus, in the high-temperature limit the experimentally observed heat capacity $C_p$ grows linearly with increasing the temperature, but $C_V$ tends to a constant value. Equation (2) allows one to find the isochoric heat capacity using the measured isobaric heat capacity and the quantities $n, \Gamma, \gamma_T$.

It turns out that at high temperature the heat capacity $C_V$ contains a linear in the temperature correction to the Dulong-Petit law $C_V = 3N$. This deviation can be explained by taking into account the interaction of phonons. Note that an early attempt to explain the high temperature deviation in the behavior of the diamond heat capacity from the standard Dulong – Petit law was made in [11]. The proposed therein explanation is based on the choice of the anharmonic potential of atomic interaction, in the same way as in a diatomic molecule.

For the most crystals, the isochoric heat capacity $C_V$ obeys the deviation from the Dulong – Petit law: it linearly decreases with increasing the temperature. In diamond-like crystals, such a deviation receives the opposite sign. In this paper we show how these features in the behavior of the high-temperature heat capacity of diamond can be described in framework of the many-particle approach.

In section 2 we briefly comment on the basic equations of [2,3]. Expression for the free energy and equation for the speed of interacting phonons are obtained there in the general form. In section 3 these expressions are calculated for the isotropic medium, and the nonlinearity parameter is introduced. Section 4 describes how to obtain the elastic moduli of the approximate isotropic medium from the elastic moduli of a crystal. Section 5 contains calculations of the high-temperature heat capacity of diamond. For comparison, same calculation is given for NaCl as a typical crystal of the cubic crystal system. Our conclusions bring together in the last section.

## 2. Self-consistent description of interacting phonons

We present relations on which the approach developed in [2,3] is based. Let phonons in the crystal lattice be described by the Hamiltonian density operator

$$\mathrm{H}(\mathbf{r}) = \frac{\pi_a(\mathbf{r})^2}{2\rho} + U_2(\mathbf{r}) + U_3(\mathbf{r}) + U_4(\mathbf{r}), \qquad (3)$$

where quadratic, cubic, and fourth-order in the strain tensor $u_{ij} = \frac{1}{2}(\nabla_j u_i + \nabla_i u_j + \nabla_i u_a \nabla_j u_a)$ terms have the form

$$U_2 = \frac{1}{2}\lambda_{aibj} u_{ai} u_{bj}, \quad U_3 = \frac{1}{6}\lambda_{aibjck} u_{ai} u_{bj} u_{ck}, \quad U_4 = \frac{1}{24}\lambda_{aibjckdl} u_{ai} u_{bj} u_{ck} u_{dl}, \qquad (4)$$

and $\rho$ is the density, $u_i(\mathbf{r})$ is the displacement vector field, $\pi_a(\mathbf{r}) = \rho \dot{u}_a(\mathbf{r})$ is the canonical momentum. Here and further on, the standard agreement on the summation over repeated indices is used. Due to the symmetry of the strain tensor $u_{ij} = u_{ji}$, the elastic moduli satisfy the known symmetry conditions for both permutations of pairs of indices and permutations of indices within the each pair [12]. With the fourth order in the displacement vector gradients accuracy, the Hamiltonian density takes the form

$$\mathrm{H}(\mathbf{r}) = \frac{\pi_a(\mathbf{r})^2}{2\rho} + \frac{1}{2}\lambda_{aibj} \nabla_i u_a \nabla_j u_b + \tilde{U}_3 + \tilde{U}_4, \qquad (5)$$

where

$$\tilde{U}_3 = \frac{1}{2}\lambda_{aibj} \nabla_i u_a \nabla_j u_c \nabla_b u_c + \frac{1}{6}\lambda_{aibjck} \nabla_i u_a \nabla_j u_b \nabla_k u_c,$$

$$\tilde{U}_4 = \frac{1}{8}\lambda_{aibj} \nabla_a u_c \nabla_i u_c \nabla_b u_s \nabla_j u_s + \frac{1}{4}\lambda_{aibjck} \nabla_i u_a \nabla_j u_b \nabla_k u_s \nabla_c u_s + \frac{1}{24}\lambda_{aibjckdl} \nabla_i u_a \nabla_j u_b \nabla_k u_c \nabla_l u_d. \qquad (6)$$

In the quantum description, the displacement and the canonical momentum should be considered as operators for which the well-known commutation relations are valid:

$$\pi_a(\mathbf{r}) u_b(\mathbf{r}') - u_b(\mathbf{r}') \pi_a(\mathbf{r}) = -i\hbar \delta_{ab} \delta(\mathbf{r} - \mathbf{r}'),$$
$$u_a(\mathbf{r}) u_b(\mathbf{r}') - u_b(\mathbf{r}') u_a(\mathbf{r}) = 0, \quad \pi_a(\mathbf{r}) \pi_b(\mathbf{r}') - \pi_b(\mathbf{r}') \pi_a(\mathbf{r}) = 0. \qquad (7)$$

In general the elastic moduli are functions of the temperature. Here as well as in the standard Debye model, we will neglect this effect, as well as the difference between the isothermal and adiabatic modules. In what follows we will take into account only the temperature dependence of the observed values, associated with the phonon excitation. The total Hamiltonian $H = \int \mathrm{H}(\mathbf{r})d\mathbf{r}$ is the sum of the free phonon Hamiltonians and their interactions $H = H_0 + H_I$, where

$$H_0 = \int \left[\frac{\pi_a(\mathbf{r})^2}{2\rho} + \frac{1}{2}\lambda_{aibj}\nabla_i u_a \nabla_j u_b\right]d\mathbf{r}, \qquad H_I = \int \left[\tilde{U}_3(\mathbf{r}) + \tilde{U}_4(\mathbf{r})\right]d\mathbf{r}. \qquad (8)$$

To describe the system of interacting phonons, we use the self-consistent field method in the formulation developed for fermionic [13] and for bosonic [14, 15] systems. The implementation of this method on the example of the anharmonic oscillator was demonstrated in [16, 17]. In the proposed formulation, the method of a self-consistent field is introduced at the level of the Hamiltonian, rather than at the level of the equations of motion. This makes possible to achieve the fulfillment of all thermodynamic relations. In this method, the total Hamiltonian is represented as the sum of two terms

$$H = H_S + H_C. \qquad (9)$$

Here, the approximating self-consistent Hamiltonian

$$H_S = \int \left[\frac{\pi_a^2}{2\rho} + \frac{\tilde{\lambda}}{2}\nabla_i u_a \nabla_i u_a\right]d\mathbf{r} + \varepsilon_0 \qquad (10)$$

describes "free" phonons with renormalized speed, and the correlation Hamiltonian

$$H_C = \int \left[\frac{1}{2}\left(\lambda_{aibj} - \tilde{\lambda}\delta_{ij}\delta_{ab}\right)\nabla_i u_a \nabla_j u_b + \tilde{U}_3 + \tilde{U}_4\right]d\mathbf{r} - \varepsilon_0 \qquad (11)$$

describes the interaction of these phonons. The self-consistent Hamiltonian (10) contains the only one effective modulus of elasticity $\tilde{\lambda}$ and describes the phonon system in the isotropic approximation when phonons with arbitrary polarization have the same speed. In addition, $H_S$ includes $\varepsilon_0$, which does not contain operators. The inclusion of this term is essential, since it describes the change in the ground state during the transition from the exact Hamiltonian to the self-consistent one. Thus, by means of the renormalization of the elastic modulus, the main interaction between the original phonons in the isotropic approximation is taken into account in Hamiltonian (10), and Hamiltonian (11) describes the residual interaction, which is not included in the self-consistent field method. The self-consistent Hamiltonian (10) in the representation of phonon creation and annihilation operators, which satisfy the standard commutation relations $\left[b_{k\alpha},b^{+}_{k'\alpha'}\right]=\delta_{kk'}\delta_{\alpha\alpha'}$, $\left[b_{k\alpha},b_{k'\alpha'}\right]=\left[b^{+}_{k\alpha},b^{+}_{k'\alpha'}\right]=0$, takes the form

$$H_S = \hbar\sum_{k,\alpha}\omega(k)b^{+}_{k\alpha}b_{k\alpha} + \frac{3}{2}\hbar\sum_{k}\omega(k) + \varepsilon_0, \qquad (12)$$

where $\omega(k)=c_S k$, and the phonon speed with arbitrary polarization is $c_S = \sqrt{\tilde{\lambda}/\rho}$. The free energy of the system with Hamiltonian (12) is given by

$$F = \varepsilon_0 + \frac{3\hbar}{2}\sum_{k}\omega(k) + 3T\sum_{k}\ln\left(1-e^{-\beta\hbar\omega(k)}\right). \qquad (13)$$

The value $\varepsilon_0$ is found from the condition $\langle H\rangle = \langle H_S\rangle$, where the average is carried out with the statistical operator $\hat{\rho}=\exp\beta(F-H_S)$, $\beta=1/T$ is the inverse temperature. The result is [3]

$$\varepsilon_0 = \frac{3\hbar}{2c_S}\left[\frac{1}{3\rho}\sum_{k}\frac{k_i k_j \lambda_{aiaj}}{k}\left(f_k+\frac{1}{2}\right) - c_S^2\sum_{k}k\left(f_k+\frac{1}{2}\right)\right] + \frac{\hbar^2}{8V\rho^2 c_S^2}I, \qquad (14)$$

where

$$I \equiv \sum_{k_1,k_2}\frac{(f_{k_1}+1/2)(f_{k_2}+1/2)}{k_1 k_2}\{\lambda_{aiajbkbl}k_{1i}k_{1j}k_{2k}k_{2l} + \\ +2\lambda_{aiajck}\left[3k_{1i}k_{1j}k_{2k}k_{2c}+2k_{1i}k_{1k}k_{2j}k_{2c}\right] + 3\lambda_{aibj}\left[3k_{1i}k_{1a}k_{2b}k_{2j}+2k_{1i}k_{1b}k_{2a}k_{2j}\right]\}, \qquad (15)$$

and $f_k = \left[e^{\beta\hbar\omega(k)}-1\right]^{-1}$ is the phonon distribution function. The phonon speed $c_S$, renormalized due to the interaction, can be found from the requirement of the free energy (13) minimum, $\partial F/\partial c_S = 0$. As a result, we obtain a nonlinear equation, which determines the speed of "new" phonons

$$c_S^2 = \frac{2\pi^2}{3\rho V}J^{-1}\sum_{k}\frac{k_i k_j \lambda_{aiaj}}{k}\left(f_k+\frac{1}{2}\right) + \frac{\hbar\pi^2}{6\rho^2 V^2 c_S}\frac{I}{J} \qquad (16)$$

where $J = \int_0^{k_D}\left(f_k+\frac{1}{2}\right)k^3 dk$, and the upper integration limit is the Debye wave number $k_D = \left(6\pi^2 N/V\right)^{1/3}$ [18]. Equation (16) is valid for the self-consistent description of the nonlinear properties of crystals of arbitrary symmetry.

In the absence of interaction and neglecting the nonlinear effects, the phonons are the same as that of the Debye model. They are naturally called "bare" or "Debye". Phonons whose speed is renormalized due to the interaction in accordance with (16) will be called the "self-consistent" ones. Even in the case of neglecting the dependence of the modules $\lambda_{aiaj}$ on the temperature, that is assumed in the ordinary

Debye model, the renormalized speed $c_S$ of our approach depends substantially on the temperature, since it is expressed in terms of integrals of the distribution function. In the considered approach, the parameter $\tilde{\lambda}$ is chosen so that Hamiltonian (10) is as close as possible to the exact Hamiltonian $H = H_0 + H_I$ and therefore describes the phonon system with the best approximation with the quadratic Hamiltonian [3].

## 3. Thermodynamic properties of nonlinear crystals in the isotropic medium approximation

Calculations of the renormalized speed (16) and of the thermodynamic properties of a certain symmetry nonlinear crystal is a complicated and cumbersome procedure that should be separately performed for each crystal class. In many cases, properties of the crystal can be described qualitatively, and even quantitatively, once anisotropy is not strong enough, in the isotropic medium approximation. In [4] such a method was developed to describe elastic waves in crystals. To account of nonlinear effects, that are cubic in the strain tensor, such a method was used in [5]. For a more consistent description of thermodynamic, and in some cases, kinetic properties of the crystal, fourth-order terms in the strain tensor should also be taken into account. The model of the isotropic medium that describes properties of a crystal allows one to substantially simplify the results and to generalize the method to crystals of arbitrary symmetry. In this case, parameters of the isotropic medium model should be found for a crystal of each symmetry from the coincidence with the exact moduli of elasticity condition.

Let us consider the interacting phonons in the isotropic medium in more detail. In this case, the second-order elastic modulus tensor is

$$\lambda_{aibj} = \lambda \delta_{ai} \delta_{bj} + \mu (ij, ba), \tag{17}$$

where $\lambda, \mu$ are the Lame coefficients. For brevity, we have used the symbol $(ij, ab) \equiv \delta_{ij}\delta_{ab} + \delta_{ia}\delta_{jb}$. The third and fourth order anharmonic elastic moduli tensors in the isotropic elastic medium have the form

$$\begin{aligned}\lambda_{aibjck} &= A\delta_{ai}\delta_{bj}\delta_{ck} + B\left[\delta_{ai}(jk, cb) + \delta_{bj}(ik, ca) + \delta_{ck}(ij, ba)\right] + \\ &+ C\left[\delta_{ac}(ij, bk) + \delta_{ak}(ij, bc) + \delta_{ic}(jk, ab) + \delta_{ik}(ab, jc)\right],\end{aligned} \tag{18}$$

$$\lambda_{aibjckdl} = C_1 \delta_{ai}\delta_{bj}\delta_{ck}\delta_{dl} + C_2 \lambda^{(2)}_{aibjckdl} + C_3 \lambda^{(3)}_{aibjckdl} + C_4 \lambda^{(4)}_{aibjckdl} + C_5 \lambda^{(5)}_{aibjckdl}, \tag{19}$$

where

$$\begin{aligned}\lambda^{(2)}_{aibjckdl} &\equiv \delta_{ai}\delta_{bj}(lk, cd) + \delta_{ai}\delta_{ck}(jl, db) + \delta_{ai}\delta_{dl}(jk, cb) + \\ &+ \delta_{bj}\delta_{ck}(il, da) + \delta_{bj}\delta_{dl}(ik, ca) + \delta_{ck}\delta_{dl}(ij, ba),\end{aligned} \tag{20}$$

$$\lambda^{(3)}_{aibjckdl} \equiv \delta_{ai}\left[\delta_{bd}(jk,cl)+\delta_{bl}(jk,cd)+\delta_{jd}(kl,bc)+\delta_{jl}(bc,kd)\right]+$$
$$+\delta_{bj}\left[\delta_{ac}(kl,di)+\delta_{ci}(kl,da)+\delta_{ka}(li,cd)+\delta_{ik}(ad,lc)\right]+ \quad (21)$$
$$+\delta_{ck}\left[\delta_{bd}(il,ja)+\delta_{dj}(il,ba)+\delta_{lb}(ij,da)+\delta_{jl}(ab,di)\right]+$$
$$+\delta_{dl}\left[\delta_{ac}(ij,bk)+\delta_{ak}(ij,bc)+\delta_{ic}(jk,ab)+\delta_{ik}(ab,jc)\right],$$

$$\lambda^{(4)}_{aibjckdl} \equiv (il,da)(jk,cb)+(ik,ca)(jl,db)+(ij,ba)(kl,dc), \quad (22)$$

$$\lambda^{(5)}_{aibjckdl} \equiv (ab,ci)(jl,dk)+(ij,ca)(kl,db)+(ik,ba)(jl,dc)+(ij,ka)(bl,dc)+$$
$$+(ab,jc)(il,dk)+(ij,bc)(kl,da)+(ab,jk)(il,dc)+(ij,bk)(cl,da)+ \quad (23)$$
$$+(ad,ci)(jk,lb)+(jl,cb)(ik,da)+(jk,db)(il,ca)+(il,ka)(cj,bd).$$

The sixth-order tensor (18) is determined by three elastic moduli $A, B, C$, and the eighth-order tensor (19) is determined by five elastic moduli $C_1 \div C_5$. When taking into account effects of the order not higher than four, four invariants can be constructed from the strain tensor:

$$J_1 = u_{ii}, \quad J_2 = u_{ai}u_{ia}, \quad J_3 = u_{ai}u_{ib}u_{ba}, \quad J_4 = u_{ai}u_{ib}u_{bj}u_{ja}. \quad (24)$$

These invariants are not independent due to the relation

$$J_4 = \frac{1}{6}J_1^4 - J_1^2 J_2 + \frac{4}{3}J_1 J_3 + \frac{1}{2}J_2^2. \quad (25)$$

The contribution to the free energy of the third and fourth order terms in the strain tensor is given by equations

$$F_3 = \frac{1}{3!}\left(AJ_1^3 + 5BJ_1J_2 + 8CJ_3\right), \quad (26)$$

$$F_4 = \frac{1}{4!}\left(C_1 J_1^4 + 12C_2 J_1^2 J_2 + 32C_3 J_1 J_3 + 12C_4 J_2^2 + 48C_5 J_4\right). \quad (27)$$

Taking into account the relation (25), equation (27) can be written in the form

$$F_4 = \frac{1}{4!}\left(DJ_1^4 + 12EJ_1^2 J_2 + 32FJ_1 J_3 + 12GJ_2^2\right), \quad (28)$$

where

$$D = C_1 + 8C_5, \quad E = C_2 - 4C_5, \quad F = C_3 + 2C_5, \quad G = C_4 + 2C_5. \quad (29)$$

Thus, there are two elastic moduli in the linear theory $\lambda, \mu$, three third order modules $A, B, C$ and four fourth order modules $D, E, F, G$ [6–9]. In the considered case, the value (15) is determined in terms of the elastic moduli as

$$I \equiv \sum_{k_1,k_2} \frac{(f_{k_1}+1/2)(f_{k_2}+1/2)}{k_1 k_2}\left[V_0 k_1^2 k_2^2 + V_1(\mathbf{k}_1\mathbf{k}_2)^2\right], \tag{30}$$

where

$$\begin{aligned} V_0 &= 9\lambda + 6\mu + 6A + 32B + 32C + D + 8E + 8F + 18G, \\ V_1 &= 6\lambda + 24\mu + 4A + 48B + 88C + 8E + 40F + 10G. \end{aligned} \tag{31}$$

Taking into account these relations, after integrating (30) over the angles, we arrive at the following equation, which determines the speed of self-consistent phonons in the isotropic elastic medium

$$c_S^2 = c_0^2 + \frac{\hbar}{24\pi^2 \rho^2 c_S}\left(V_0 + \frac{V_1}{3}\right) J, \tag{32}$$

where the average speed of the "bare" phonons is defined by

$$c_0^2 = \frac{1}{3}\left(2c_t^2 + c_l^2\right) = \frac{(\lambda+4\mu)}{3\rho}, \tag{33}$$

and the longitudinal and transverse velocities of sound are determined by the known relations $c_l^2 = (\lambda+2\mu)/\rho$, $c_t^2 = \mu/\rho$ [12]. Note that the definition (33) differs from the definition of the average Debye rate [1]:

$$\frac{1}{c_D^3} = \rho^{3/2}\left[\frac{2}{\mu^{3/2}} + \frac{1}{(\lambda+2\mu)^{3/2}}\right]. \tag{34}$$

In the model, where the average speed or the Debye energy are phenomenological parameters, this difference in the definition is not significant. However, since the average velocities (33) and (34) are expressed differently through the elastic moduli, this distinction should be taken into account in a more exact description. The standard Debye energy is determined by the relation $\Theta_D \equiv \hbar c_D k_D$ [18], but in the considered approach, as it can be seen from equation (32), the natural definition of the Debye energy is $\Theta_0 \equiv \hbar c_0 k_D$. Since now, besides the speed of the "bare" phonons $c_0$, there arises the speed of the self-consistent phonons $c_S$, so it is natural to define the "self-consistent Debye energy" $\tilde{\Theta}_D \equiv \hbar c_S k_D$, which, unlike the standard definition of the Debye energy $\Theta_D$ or $\Theta_0$, is a function of the temperature. Note, however, that although in its original formulation the Debye energy is assumed to be independent on the temperature, in practice experiments (e.g. [19–21]) reveal the temperature dependence of the Debye energy. In the considered approach, the self-consistent Debye energy $\tilde{\Theta}_D$ significantly depends on the temperature due to the phonon-phonon interaction already at the starting point and, therefore, this approach better reflects the real situation. Although the shortcomings of the Debye model, associated with a very simplified choice of the spectral density which does not take into account the details of the structure of the lattice, remain also in our approach.

For the further reference, it is convenient to introduce the symbol $\sigma$ for the ratio of the renormalized due to the phonon-phonon interaction sound speed $c_S$ to the original average sound speed $c_0$, or, which is the same, ratio of the self-consistent Debye energy to the standard one:

$$\sigma \equiv c_s/c_0 = \tilde{\Theta}_D/\Theta_0 . \tag{35}$$

Equation (32), on account of the introduced quantity (35), can be written in the dimensionless form [3,4]

$$(\sigma^2 - 1)\sigma = \Lambda \Phi\left(\frac{\sigma}{\tau}\right), \tag{36}$$

where $\tau \equiv T/\Theta_0$ is the dimensionless temperature. We have taken into account that $J = \frac{k_D^4}{8}\Phi\left(\frac{\sigma}{\tau}\right)$, where $\Phi(x) \equiv 1 + \frac{8}{3x}D(x)$, $D(x) = \frac{3}{x^3}\int_0^x \frac{z^3 dz}{e^z - 1}$ is the Debye function. Equation (36) contains a single dimensionless parameter that characterizes the system:

$$\Lambda \equiv \frac{\Theta_0}{32\rho M c_0^4}\left(V_0 + \frac{V_1}{3}\right), \tag{37}$$

where $M$ is the mass of the lattice atom. In general, there are no restrictions on the sign of this parameter. However, calculations show that for the most substances the sign of $\Lambda$ is positive. This leads to the fact that in most cases the phonon speed $c_S$ and the self-consistent Debye energy $\tilde{\Theta}_D$ increase with the temperature [2, 3]. For diamond and crystals with the similar structure, such as germanium and silicon, this parameter turns out to be negative. Further on we will study the solely diamond crystal case.

## 4. Approximation of nonlinear characteristics of crystals by the elastic moduli of the isotropic medium

A large number of elastic moduli describes the nonlinear elastic properties of crystals of different symmetry. Most simply is to describe nonlinear properties of the isotropic medium, where there are nine modules upon the expansion of the free energy up to the fourth power in the strain tensor. In most cases, crystals can also be well described using the isotropic medium model, if its parameters are optimally selected. In [4] it was proposed to choose the elastic moduli of the approximating linearized isotropic medium from the requirement of the minimum of the quantity

$$I_2 \equiv \left|\lambda_{aibj} - \lambda_{aibj}^{(0)}\right|^2, \tag{38}$$

where $\lambda_{aibj}$ is the modulus of elasticity of a crystal, $\lambda_{aibj}^{(0)}$ is the elastic modulus of the isotropic medium. As it was shown in [22], the free energy of the isotropic medium turns out to be as close as possible to the free energy of the crystal in the case. It is natural for the approximation of nonlinear elastic moduli to use the analogous condition (38) [5], minimizing the quantities

$$I_3 \equiv \left|\lambda_{aibjck} - \lambda_{aibjck}^{(0)}\right|^2, \qquad I_4 \equiv \left|\lambda_{aibjckdl} - \lambda_{aibjckdl}^{(0)}\right|^2, \qquad (39)$$

where $\lambda_{aibjck}$, $\lambda_{aibjckdl}$ are the elastic moduli tensors of the crystal, $\lambda_{aibjck}^{(0)}$, $\lambda_{aibjckdl}^{(0)}$ are the elastic moduli tensors of the isotropic medium. Taking the derivatives of $I_2, I_3, I_4$ with respect to the elastic moduli of the isotropic medium and setting them to zero, we obtain the systems of equations whose solutions are the elastic moduli of a reduced isotropic medium, expressed in terms of the following convolutions of the elastic tensors of a real crystal:

$$\lambda_1^{(2)} \equiv \lambda_{iikk}, \quad \lambda_2^{(2)} \equiv \lambda_{ikik},$$
$$\lambda_1^{(3)} \equiv \lambda_{iikkll}, \quad \lambda_2^{(3)} \equiv \lambda_{ikklli}, \quad \lambda_3^{(3)} \equiv \lambda_{iikllk}, \qquad (40)$$
$$\lambda_1^{(4)} \equiv \lambda_{iikkllpp}, \quad \lambda_2^{(4)} \equiv \lambda_{iikklppl}, \quad \lambda_3^{(4)} \equiv \lambda_{iikllppk}, \quad \lambda_4^{(4)} \equiv \lambda_{ikkilppl}, \quad \lambda_5^{(4)} \equiv \lambda_{ikkllppi}.$$

In terms of these quantities, the reduced elastic moduli are

$$\lambda = \frac{1}{15}\left(2\lambda_1^{(2)} - \lambda_2^{(2)}\right), \qquad \mu = \frac{1}{30}\left(-\lambda_1^{(2)} + 3\lambda_2^{(2)}\right), \qquad (41)$$

$$A = \frac{1}{105}\left(8\lambda_1^{(3)} + 8\lambda_2^{(3)} - 15\lambda_3^{(3)}\right), \quad B = \frac{1}{210}\left(-15\lambda_1^{(3)} - 12\lambda_2^{(3)} + 19\lambda_3^{(3)}\right),$$
$$C = \frac{1}{210}\left(2\lambda_1^{(3)} + 9\lambda_2^{(3)} - 9\lambda_3^{(3)}\right), \qquad (42)$$

$$D = \frac{1}{56700}\left(1123\lambda_1^{(4)} - 618\lambda_2^{(4)} - 2296\lambda_3^{(4)} - 1671\lambda_4^{(4)} + 3522\lambda_5^{(4)}\right),$$
$$E = \frac{1}{113400}\left(-103\lambda_1^{(4)} + 438\lambda_2^{(4)} + 376\lambda_3^{(4)} + 411\lambda_4^{(4)} - 1002\lambda_5^{(4)}\right),$$
$$F = \frac{1}{226800}\left(-287\lambda_1^{(4)} + 282\lambda_2^{(4)} + 1184\lambda_3^{(4)} + 39\lambda_4^{(4)} - 978\lambda_5^{(4)}\right), \qquad (43)$$
$$G = \frac{1}{226800}\left(-557\lambda_1^{(4)} + 822\lambda_2^{(4)} + 104\lambda_3^{(4)} + 2469\lambda_4^{(4)} - 2598\lambda_5^{(4)}\right).$$

For the cubic crystals, to which diamond belongs, the following components are selected as the independent components of the elasticity tensors of the fourth, sixth and eighth ranks in the matrix representation [4,6]: 1) $c_{11}, c_{12}, c_{44}$; 2) $c_{111}, c_{112}, c_{155}, c_{123}, c_{144}, c_{456}$; 3) $c_{1111}, c_{1112}, c_{1166}, c_{1122}, c_{1266}, c_{4444}, c_{1123}, c_{1144}, c_{1244}, c_{1456}, c_{4466}$. In this

case, equations (41) - (43) for the elastic moduli of the reduced isotropic medium take the form

$$\lambda = \frac{1}{5}(c_{11} + 4c_{12} - 2c_{44}), \qquad \mu = \frac{1}{5}(c_{11} - c_{12} + 3c_{44}), \qquad (44)$$

$$A = \frac{1}{35}(c_{111} + 18c_{112} + 16c_{123} - 30c_{144} - 12c_{155} + 16c_{456}),$$

$$B = \frac{1}{35}(c_{111} + 4c_{112} - 5c_{123} + 19c_{144} + 2c_{155} - 12c_{456}), \qquad (45)$$

$$C = \frac{1}{35}(c_{111} - 3c_{112} + 2c_{123} - 9c_{144} + 9c_{155} + 9c_{456}),$$

$$D = \frac{1}{315}(c_{1111} + 32c_{1112} + 36c_{1122} + 204c_{1123} - 132c_{1144} -$$
$$-24c_{1166} - 312c_{1244} - 36c_{1266} + 240c_{1456} + 6c_{4444} + 12c_{4466}),$$

$$E = \frac{1}{315}(c_{1111} + 14c_{1112} + 9c_{1122} - 3c_{1123} + 21c_{1144} -$$
$$-6c_{1166} + 48c_{1244} - 48c_{1456} - 3c_{4444} - 6c_{4466}),$$

$$F = \frac{1}{630}(2c_{1111} + 10c_{1112} - 9c_{1122} - 24c_{1123} + 6c_{1144} + \qquad (46)$$
$$+6c_{1166} + 78c_{1244} + 36c_{1266} - 6c_{1456} - 15c_{4444} - 30c_{4466}),$$

$$G = \frac{1}{630}(2c_{1111} - 8c_{1112} + 27c_{1122} - 42c_{1123} + 96c_{1144} +$$
$$+24c_{1166} + 60c_{1244} - 54c_{1266} - 168c_{1456} + 39c_{4444} + 78c_{4466}).$$

Equations (44) - (46) are used here for computing the parameters of the proposed model for diamond crystals. The values of the independent components of the elasticity tensors of the fourth, sixth, and eighth ranks, which are given in Tables 1 and 2, are either measured experimentally or calculated theoretically [20,21]. As it has been noted, when the approximate description is used, the temperature dependence of the elastic moduli and the difference between the isothermal and adiabatic modules could be ignored.

**Table 1. Independent components of the fourth and sixth elasticity tensors of diamond ($10^{11}$ dyn cm$^{-2}$)**

| $C_{11}$ | $C_{12}$ | $C_{44}$ | $C_{111}$ | $C_{112}$ | $C_{155}$ | $C_{123}$ | $C_{144}$ | $C_{456}$ |
|---|---|---|---|---|---|---|---|---|
| 108 | 12.4 | 57.8 | -761 | -226 | -280 | 210 | -178 | -82.0 |

**Table 2. Independent components of the eighth rank elastic tensor of diamond ($10^{11}$ dyn cm$^{-2}$)**

| $C_{1111}$ | $C_{1112}$ | $C_{1166}$ | $C_{1122}$ | $C_{1266}$ | $C_{4444}$ | $C_{1123}$ | $C_{1144}$ | $C_{1244}$ | $C_{1456}$ | $C_{4466}$ |
|---|---|---|---|---|---|---|---|---|---|---|
| 2669 | 946 | 1074 | 607 | 819 | 1132 | -42.5 | -138.5 | -26.4 | 48.7 | 52.8 |

Using the data given in Tables 1, 2, equations (44) - (46) result in the elastic coefficients of the reduced isotropic medium for diamond (Table 3).

**Table 3. The nine reduced moduli of elasticity of diamond ($10^{11}$ dyn cm$^{-2}$)**

| $\lambda$ | $\mu$ | A | B | C | D | E | F | G |
|---|---|---|---|---|---|---|---|---|
| 8.4 | 53.8 | 169.1 | -162.1 | -37.7 | 637.5 | 186.8 | -280.3 | 183.9 |

Using these data together with values of the density $\rho$ and of the atomic mass $M$ makes possible to calculate the parameters of the generalized Debye theory for the self-consistent phonons [2,3] in diamond, which are listed in Table 4.

**Table 4. Values of the parameters $V_0$, $V_1$ (31) и $\Lambda$ (37), which determine the nonlinear properties of diamond crystal.**

| $\rho$, g/cm$^3$ | M, $10^{-23}$ g | $V_0$, $10^{11}$dyn/cm$^2$ | $V_1$, $10^{11}$dyn/cm$^2$ | $\Lambda$ |
|---|---|---|---|---|
| 3.5 | 1.99 | -1784.98 | -2030.2 | -0.00817 |

The values of the temperature independent longitudinal and transverse sound velocities, the average velocities of (33), (34) and, the corresponding Debye temperatures, are given in Table 5.

**Table 5. Temperature independent longitudinal, transverse and average sound speeds and the Debye temperature of diamond.**

| $c_t$ $10^5$ sm/c | $c_l$ $10^5$ sm/c | $c_D$ $10^5$ sm/c | $c_0$ $10^5$ sm/c | $k_D$ $10^8$ cm$^{-1}$ | $\Theta_D$ K | $\Theta_0$ K |
|---|---|---|---|---|---|---|
| 12.31 | 18.09 | 9.37 | 14.59 | 2.18 | 1560.6 | 2430.0 |

## 5. The temperature dependences of the Debye energy. Heat capacity

Account of the phonon-phonon interaction leads to the appearance of the dependence of the average phonon speed on the temperature (32), (36) even if such a dependence was absent in the linear approximation. If we determine the Debye energy or the Debye temperature through this speed $\tilde{\Theta}_D \equiv \hbar c_s k_D$, the Debye energy will also depend on the temperature. Figure 1 shows the temperature dependence of the self-consistent phonon velocity and the Debye energy of diamond, calculated with the nonlinearity parameter $\Lambda$ given in Table 4.

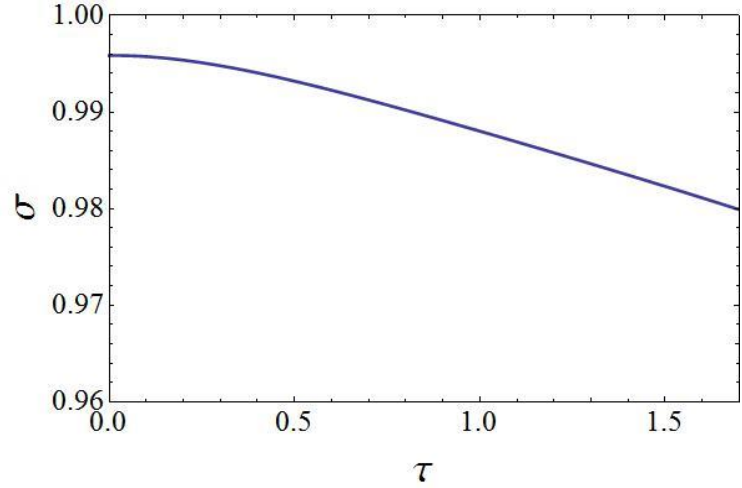

**Fig. 1. Dependence of the self-consistent phonon speed and the Debye energy of diamond $\sigma \equiv c_s/c_0 = \tilde{\Theta}_D/\Theta_0$ on the temperature $\tau \equiv T/\Theta_0$.**

For most solids the nonlinearity parameter $\Lambda$ is positive, and the average speed of the self-consistent phonons increases with the temperature [3]. For diamond and some other diamond-like crystals, the parameter $\Lambda$ turned out to be negative, so that the phonon speed and the Debye energy decrease with the temperature for them (Fig.1).

Entropy can be obtained by the well-known expression $S = -(\partial F/\partial T)_V$ with the free energy (13), (14). Computing the derivative, the speed $c_s$, in view of $\partial F/\partial c_s = 0$, could be treated as a constant [2,3], then:

$$S = -N\left[3\ln\left(1-e^{-\frac{\sigma}{\tau}}\right) - 4D\left(\frac{\sigma}{\tau}\right)\right]. \tag{47}$$

From the expression for entropy (47) it follows the expression for the isochoric heat capacity $C_V = T(\partial S/\partial T)_V$ [2,3]:

$$C_V = 3N\left[4\frac{\tau}{\sigma}D\left(\frac{\sigma}{\tau}\right) - \frac{3}{e^{\sigma/\tau}-1}\right]\left(\frac{\sigma}{\tau} - \frac{d\sigma}{d\tau}\right). \tag{48}$$

The temperature derivative in (48) can be found from equation (36) [3]. As it is known, in the Debye model there is the law of the corresponding states, consisting in the fact that the heat capacity is a function of the dimensionless temperature $\tau = T/\Theta$ [1]. Account of the phonon-phonon interaction leads to the violation of this law, and every specific phonon system is additionally characterized by its dimensionless parameter $\Lambda$. The calculation of the parameter $\Lambda$ shows that it is positive for the most substances. Consequently, we get increasing the self-consistent phonon speed with the temperature and the linear in the temperature deviation of the heat capacity $C_V$ from the Dulong-Petit law [3]. Since experimentally it is usually measured the heat capacity $C_p$, which linearly grows at high temperatures (see eq. (2)), to identify the decrease of $C_V$ experimentally, it is necessary to use the equation (2) upon processing the experimental data.

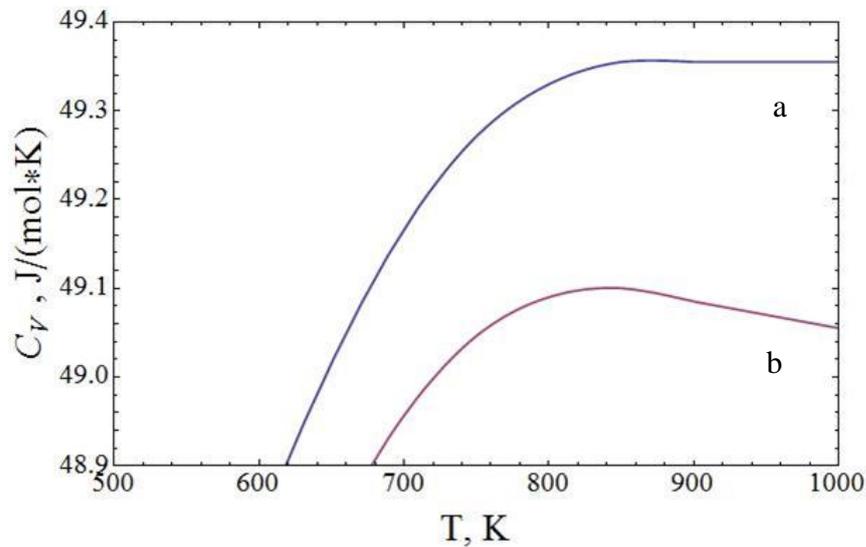

**Fig. 2. The molar heat capacity of NaCl at a constant volume: (a) constructed from the experimental data; (b) subjected to equation (2).**

Figure 2 shows, using NaCl as an example, the typical behavior of the isochoric heat capacity at high temperatures, characterizing the most substances. The experimental data are taken from [23].

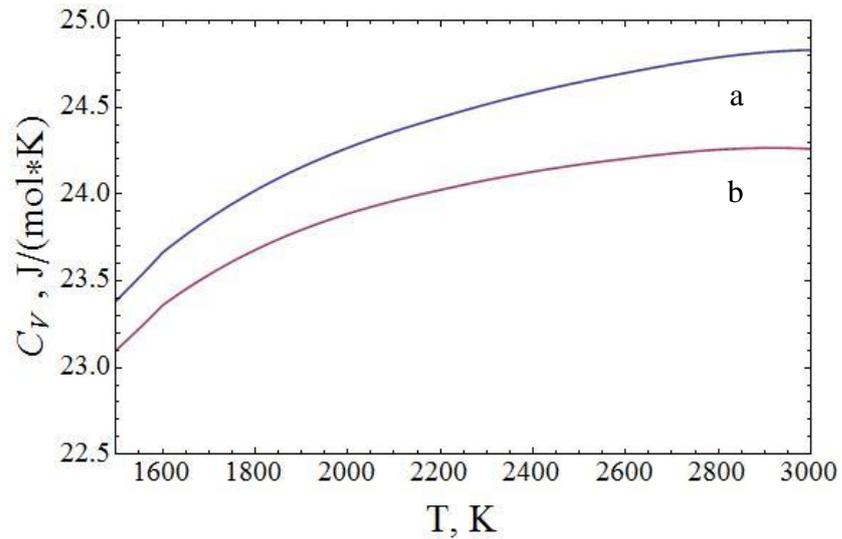

**Fig. 3. The molar heat capacity of diamond at a constant volume: (a) constructed from the experimental data; (b) calculated by equation (2).**

Figure 3 shows the temperature dependence of the isochoric heat capacity of diamond, plotted by use of the experimental data from [24, 25]. As we can see, the behavior $C_v$ of diamond at high temperatures is qualitatively different from the behavior of this quantity for the most crystals of the cubic crystal system (Fig. 2).

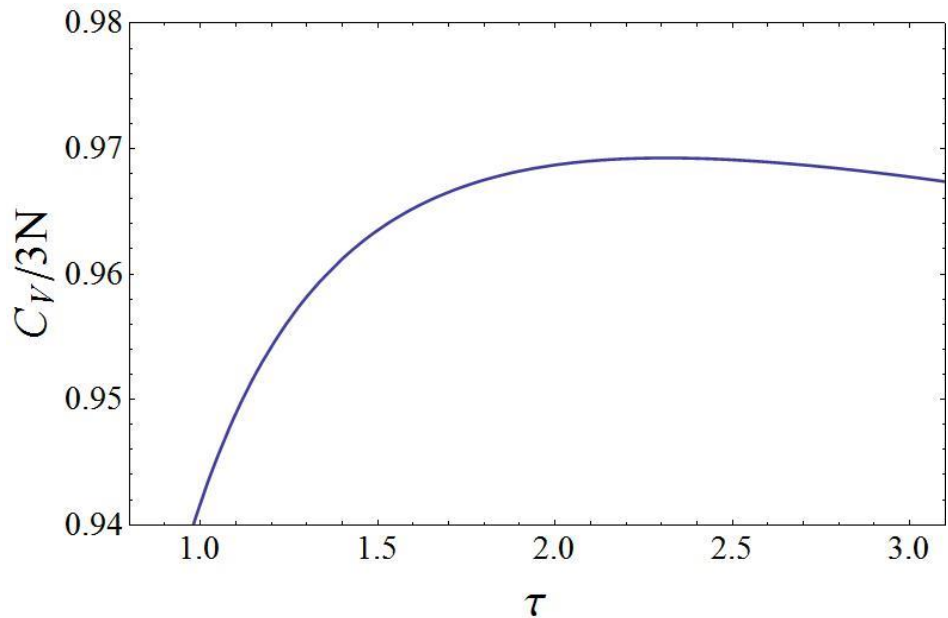

**Fig. 4. The normalized heat capacity of NaCl at a constant volume, calculated from equation (48), ($\Lambda = 0.007$).**

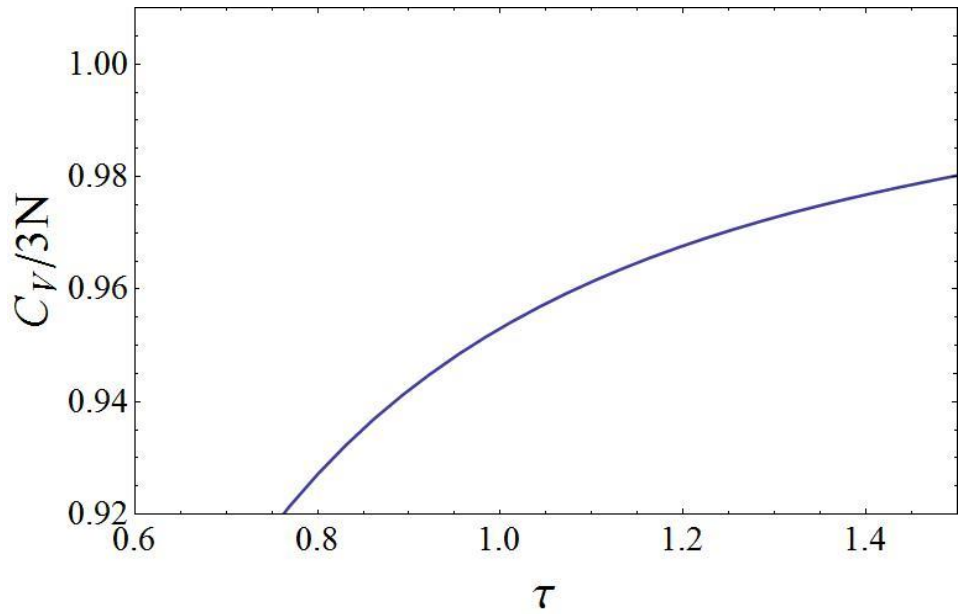

**Fig. 5. The normalized heat capacity of diamond at a constant volume, calculated from equation (48), ($\Lambda = -0.00817$).** The upper temperature limit is constrained by the melting point.

## 6. Conclusions

For the most substances, the linear in the temperature decrease of $C_v$ is observed with increasing the temperature. For crystals with diamond cubic crystal structure, this deviation occurs in the direction of increasing the heat capacity. In the framework of the approach of the self-consistent phonons, using the approximation of the elastic properties of crystals by the reduced isotropic medium, it is possible to clarify formulae for the isochoric heat capacity and the Debye energy.

Account of the phonon-phonon interaction leads to the redefinition of the phonon's speed and of the Debye energy. Their dependence on the temperature occurs. The isochoric heat capacity is no longer a constant. At low temperatures, the corrections to $C_v$ are insignificant (the phonon gas can be considered as ideal). At high temperatures, the sign of the correction depends on the sign of the non-linearity parameter $\Lambda$. It is convenient to calculate it for the isotropic medium (see eq. (37), (31)). Calculations show that for all crystals of the cubic system $\Lambda$ is positive, except of diamond and crystals with diamond structure. Figures 4, 5 show the difference in the behavior of $C_v$ for diamond and NaCl (as an example of crystals of cubic crystal system).

Using the developed approach, one can calculate corrections to the thermodynamic coefficients of crystals under conditions when the phonon gas cannot be considered as weakly interacting. Also, it is possible to obtain the temperature dependences of the longitudinal and transverse Debye energies, if we do not average the speed of phonon. Estimates show that for diamond the difference between the longitudinal and transverse Debye energies can be significant.